# Strategic tradeoffs in competitor dynamics on adaptive networks

Laurent Hébert-Dufresne,[1, 2] Antoine Allard,[3] Pierre-André Noël,[4] Jean-Gabriel Young,[5] and Eric Libby[1, *]

[1]*Santa Fe Institute, Santa Fe, NM 87501, USA*
[2]*Institute for Disease Modeling, Bellevue, WA, 98005, USA*
[3]*Centre de Recerca Matemàtica, E-08193 Bellaterra (Barcelona), Spain*
[4]*University of California, Davis CA 95616, USA*
[5]*Département de physique, de génie physique et d'optique,
Université Laval, Québec (Qc), G1V 0A6, Canada*

Recent empirical work highlights the heterogeneity of social competitions such as political campaigns: proponents of some ideologies seek debate and conversation, others create echo chambers. While symmetric and static network structure is typically used as a substrate to study such competitor dynamics, network structure can instead be interpreted as a signature of the competitor strategies, yielding competition dynamics on adaptive networks. Here we demonstrate that tradeoffs between aggressiveness and defensiveness (i.e., targeting adversaries vs. targeting like-minded individuals) creates paradoxical behaviour such as non-transitive dynamics. And while there is an optimal strategy in a two competitor system, three competitor systems have no such solution; the introduction of extreme strategies can easily affect the outcome of a competition, even if the extreme strategies have no chance of winning. Not only are these results reminiscent of classic paradoxical results from evolutionary game theory, but the structure of social networks created by our model can be mapped to particular forms of payoff matrices. Consequently, social structure can act as a measurable metric for social games which in turn allows us to provide a game theoretical perspective on online political debates.

## INTRODUCTION

Fixed resources drive competition and non-linear dynamics in socio-biological systems[1–8]. As entities compete over resources, they often face strategic decisions: pursuing one resource means foregoing another. The importance of such strategic decisions is exacerbated when resources are heterogeneous because some are ultimately more valuable than others. Many real world scenarios feature heterogeneous resources where strategic decisions determine the winner of the competition. For example, consider political campaigns, a canonical example of social competition where voters identify with one of many candidates and either try to change or reinforce the opinions of other voters. A typical strategic decision is how much time to spend debating with adversaries so as to change their opinions versus agreeing with like-minded voters. Recent studies of online conversations provide unique insights into this process [9–12]. Barberá *et al.* studied 150 million tweets on Twitter to determine how often online political discussions were debates as opposed to echo chambers where like-minded people voice a shared opinion [12]. Their results, reproduced on Fig. 1, demonstrate how users with different ideologies behave in characteristic manners. More specifically, they found that users identified as liberals are more likely to initiate cross-ideological conversations on political issues than users identified as conservatives. Similar differences in strategy between people of different ideologies have also been observed in other online discussion forums [9] including user comments on online newspapers [11]. Motivated by these examples, we consider a general model of competition between different strategies.

Classical models of competitor dynamics on networks, such as the voter model (VM) [13] and the analogous Moran process model (MP) [14], do not distinguish between resources in a way that permits consideration of strategic tradeoffs. Indeed, each competitor in an MP is defined by a single parameter that expresses their ability to indiscriminately obtain available resources. Similarly, a VM typically considers a fixed symmetric social structure (i.e. with undirected interactions) and the influence of a voter over its neighbours does not depend on their current state. The lack of state-dependent interactions is particularly limiting because resources will likely change hands, or states. As this happens, competitors may want to modify interactions to reclaim or protect resources. In the context of the earlier political example, when nodes change states they adopt different opinions/ideologies and thus probably change their strategies accordingly. Adaptive networks, where links (or their weights) change with the states of nodes, offer a natural way to model this plasticity. This rewiring allows strategies to determine both which nodes interact and how they interact, depending on their states.

Using the directed stochastic block model (SBM) to encode these strategies [15], we extend the MP and VM dynamics to an adaptive network structure to study the effects of strategic decisions. We obtain general analytical solutions for the voter model dynamics and investigate specific cases with tradeoffs between aggressiveness and defensiveness (i.e., targeting adversaries vs. targeting like-minded individuals). We show that these tradeoffs yield interesting, and even paradoxical, behaviors such as long transient dynamics, sensitive dependence to initial conditions, and non-transitive dynamics. These results are reminiscent of classic voting paradoxes and are known results from evolutionary game theory; in fact, the SBM allows us to directly map the social structure created by our model to particular cases of payoff matrices in game theory. While the mapping between the two models is not exact, we observe several interesting results of well-mixed game theory in our network model. This means on the one hand that our model can provide a network perspective to game theory, and on the other hand that social network structure can potentially be used to infer the payoff structure of an equivalent evolutionary game. For example, we show how we can use our model to infer different dynamical regimes from empirical observations of activities on Twitter and interpret the resulting network structure as a signature of competitor strategies.

---

* Correspondence to: elibby@santafe.edu

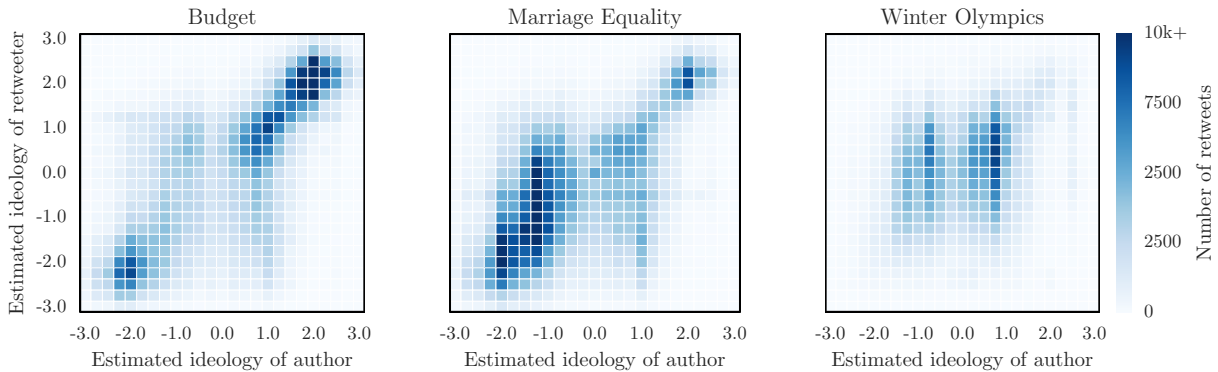

FIG. 1. **Debates and echo chambers on Twitter.** Shown is the number of retweets according to the inferred ideologies of the author and of the retweeter. The ideology of authors is estimated through their connections and ranges from strongly liberal (-3) to strongly conservative (+3). Political subjects such as the federal budget and marriage equality in the USA (left and middle panels respectively) lead to a much stronger homophily and the creation of echo chambers, while non-political subjects such as the 2014 Winter Olympics (right panel) are not polarized. Reproduced from Ref. [12].

## RESULTS

We consider competitor dynamics on an adaptive network where nodes are units of resource adopting a state according to which of the $g$ competitors currently claims them. The interactions between the nodes depend on their states and are prescribed by the $g \times g$ matrix $\boldsymbol{P}$ whose elements $p_{ij}$ correspond to the probability that a directed link exists from a node of state $i$ towards a node of state $j$. As in the traditional voter model, at every time step a randomly chosen node adopts the state of a node at the end of a randomly chosen incoming link. In this adaptive version, once a node changes state its incoming and outgoing edges are redrawn according to $\boldsymbol{P}$ reflecting its new state. Thus, the network evolves throughout the competition as nodes change hands. Other versions of adaptive coevolution of structure and voter dynamics exist [16–18] but, to the best of our knowledge, all involve symmetric strategies across competitors.

Here, the density matrix $\boldsymbol{P}$ is not only a description of the underlying structure at a given time, but directly reflects the different strategies of different competitors. For instance, a modular structure (i.e., larger $p_{ij}$ values on the diagonal or homophily) implies defensive or self-reinforcing strategies that try to prevent their nodes from switching to a different state. A fuzzy multipartite structure (i.e., larger $p_{ij}$ values off the diagonal or heterophily) reflects offensive strategies where individuals mostly target competitors. Similarly, a core-periphery structure [19] reflects a defensive competitor facing an offensive strategy (i.e., one row showing homophily and others rows showing heterophily). Our model can therefore lead to very different network structures arising from the interplay between strategies. In fact, the network structure is entirely specified by $\boldsymbol{P}$ which in turn is a direct parametrization of the strategies of the competitors.

Without any constraints on the density matrix $\boldsymbol{P}$, the optimal strategy for nodes belonging to competitor $i$ would be to fully target every state, i.e. $p_{ij} = 1\,\forall j$. However, to embody the key tradeoffs mentioned in the introduction, an obvious choice of constraints is $p_{ii} + p_{ij} = 1\;\forall j \neq i$, which forces competitors to choose between offense (targeting competitor-owned nodes) and defense (targeting self-owned nodes). We analyze the resulting dynamics for two competitors trying to capture a majority of nodes and find that there exists a single optimal strategy. In contrast, the presence of a third competitor results in much richer dynamics. We discover that there are four canonical types of dynamics for three competi-tors that can exhibit counterintuitive, nonlinear behaviors.

### The two competitor case

We begin our analysis of competitor dynamics on an adaptive network by considering a reduced form of the general framework where there are only two competitors who both face the same constraint $p_{i1} + p_{i2} = 1$ on their possible strategies. The matrix $\boldsymbol{P}$ thus has the structure

$$\boldsymbol{P} = \begin{pmatrix} p_1 & 1-p_1 \\ 1-p_2 & p_2 \end{pmatrix}, \quad (1)$$

where $0 \leq p_1, p_2 \leq 1$ are the parameters that define each strategy. The value of $p_i$ determines how much competitor $i$ influences itself, i.e., defending its own resources, while $1 - p_i$ is the influence on the opposing competitor. Since the total influence of any competitor is constrained, there is a trade-off between group cohesion $p_i$ (i.e., defense) and the deployment of effort to gather new nodes $1 - p_i$ (i.e., offense). The success of a competitor is measured by its frequency in the population, which we denote $x_i$. The frequencies $x_i$ range from 0 to 1, and only one of them is needed to fully specify the state of a system with 2 competitors since the other is constrained by the conserved population: $\sum_i x_i = 1$.

Using Eq. (1) and the conservation relation $x_2 = 1 - x_1$, we can describe the change in frequency of competitor 1, whose strategy is determined by $p_1$, by

$$\dot{x}_1 = \frac{x_1(1-x_1)(1-p_1)}{x_1(1-p_1) + (1-x_1)p_2} - \frac{x_1(1-x_1)(1-p_2)}{x_1 p_1 + (1-x_1)(1-p_2)}. \quad (2)$$

The first term corresponds to nodes belonging to competitor 1 trying to claim the remaining $1 - x_1$ fraction of nodes. The probability that an offense on a given node of competitor 2 is successful is given by the ratio of edges from competitor 1 $[x_1(1-p_1)]$ to the total number of incoming edges on that node [offense plus defense: $x_1(1-p_1) + (1-x_1)p_2$]. The second term correspond to nodes of competitor 1 being claimed by nodes of competitor 2, and is constructed using the same logic.

The three fixed points of Eq. (2) are

$$x_1^{*(1)} = 0, \quad x_1^{*(2)} = 1 \quad \text{and}$$
$$x_1^{*(3)} = (1-p_2)/(2-p_1-p_2). \quad (3)$$

Analyzing the stability of the fixed points, we find that there are two qualitative regimes depending on the value of $p_1 + p_2$ (see Supplementary Methods for analysis). If $p_1 + p_2 < 1$, both $x_1^{*(1)}$ and $x_1^{*(2)}$ are unstable and $x_1^{*(3)}$ is stable. This means that the competition will result in a mixed population where neither competitor goes extinct. The competitor who has the highest value of $p$ will make up the majority of the population. In contrast, if $p_1 + p_2 > 1$, then both $x_1^{*(1)}$ and $x_1^{*(2)}$ are stable and $x_1^{*(3)}$ is unstable. Thus, one competitor will always go extinct. The winner does not depend on strategy but rather the initial frequency. If $x_1(0)$ is greater than $x_1^{*(3)}$ then $x_1$ will win; if $x_1(0)$ is less than $x_1^{*(3)}$ then $x_1$ will go extinct.

Our stability analysis shows that coexistence is only possible if competitors adopt sufficiently offensive strategies, i.e. $p_1 + p_2 < 1$. We call such competitions "pairwise aggressive" because each competitor's offense overwhelms the defense of their opponent, i.e. $p_1 < 1 - p_2$ and $p_2 < 1 - p_1$. In Fig. 2, an example of a pairwise aggressive competition shows that as the two strategies compete for resources they shape the network topology into a disassortative structure: nodes of one competitor preferentially target nodes of the other competitor. In contrast, a competition in which $p_1 + p_2 > 1$ results in an assortative network. We call such competitions "pairwise defensive" because each competitor's defense is greater than the offense of their opponent. The ultimate result of a pairwise defensive competition is annihilation of one of the competitors (see also Fig. 2).

This result is surprising when we consider the underlying network architecture. When both strategies are defensive (high $p_i$ for all $i$), the assortative network has a highly modular structure which indicates poorly coupled subsystems occupied by different competitors. This would seem to promote coexistence as it is analogous to each competitor having a well-defined territory and rarely seeking to acquire outside nodes. Yet, this structure promotes extinction. This occurs due to a positive feedback mechanism: once a competitor loses a node, its opponent's territory grows because of the adaptive structure. This node is now more strongly defended because it is part of a larger module, and thus harder to recapture. This feedback destabilizes coexistence. Similarly, we might expect the well-mixed system architecture of the low $p$ strategy competition to allow for one competitor to rapidly capture all the nodes, but our results show the opposite.

Finally, we find that the two-competitor system has the optimal strategy $p = 1/2$, although the ultimate outcome may depend on the initial resources of the competitors. If both competitors start with the same resources $x_1(0) = x_2(0) = 1/2$, then the strategy $p = 1/2$ is unbeatable: it guarantees a win against all $p \neq 1/2$. Note that a competitor who values survival instead of winning by majority may prefer the strategy $p = 0$, which guarantees a nonzero equilibrium population for all nonzero initial conditions.

### Evolutionary game theory perspective

One striking aspect of the solutions shown in Eq. 3 is that their phenomenology is surprisingly reminiscent of the solutions of a 2-strategy game in an infinite and well-mixed population. Consider the classic example of the 2-player, 2-strategy prisoner's dilemma. At every step, two individuals in an infinite population are chosen at random. Each individual must choose to either collaborate or defect. If both collaborate, they are awarded a big payoff $R$ as a reward. If they both defect, they are awarded a small payoff $P$ as punishment. If one defects and one collaborates, they respectively get the temptation award $T$ and sucker punishment $S$. The game is thus fully parametrized by the following award (or payoff) matrix $\boldsymbol{A}$:

$$\boldsymbol{A} = \begin{pmatrix} R & S \\ T & P \end{pmatrix}, \qquad (4)$$

where $A_{ij}$ corresponds to the award given to a node in state $i$ interacting with a node in state $j$. State $i = 1$ corresponds to cooperation and is awarded either $R$ ($j = 1$) or $S$ ($j = 2$), and state $i = 2$ corresponds to defection and is awarded either $T$ ($j = 1$) or $P$ ($j = 2$). Following the presentation in Ref. [20], we denote the frequency of strategy $i$ in the infinite population as $y_i$. Its expected award is therefore $f_i = \sum_j y_j A_{ij}$, and the expected award of any random individual is $\alpha = \sum_i y_i f_i$. Under the assumption that the frequency of a strategy is proportional to its expected award, we can write the following mean-field dynamics:

$$\dot{y}_i = y_i (f_i - \alpha), \qquad (5)$$

whose fixed points are

$$y_1^{*(1)} = 0, \quad y_1^{*(2)} = 1 \quad \text{and}$$
$$y_1^{*(3)} = (P - S) / (R - S - T + P). \qquad (6)$$

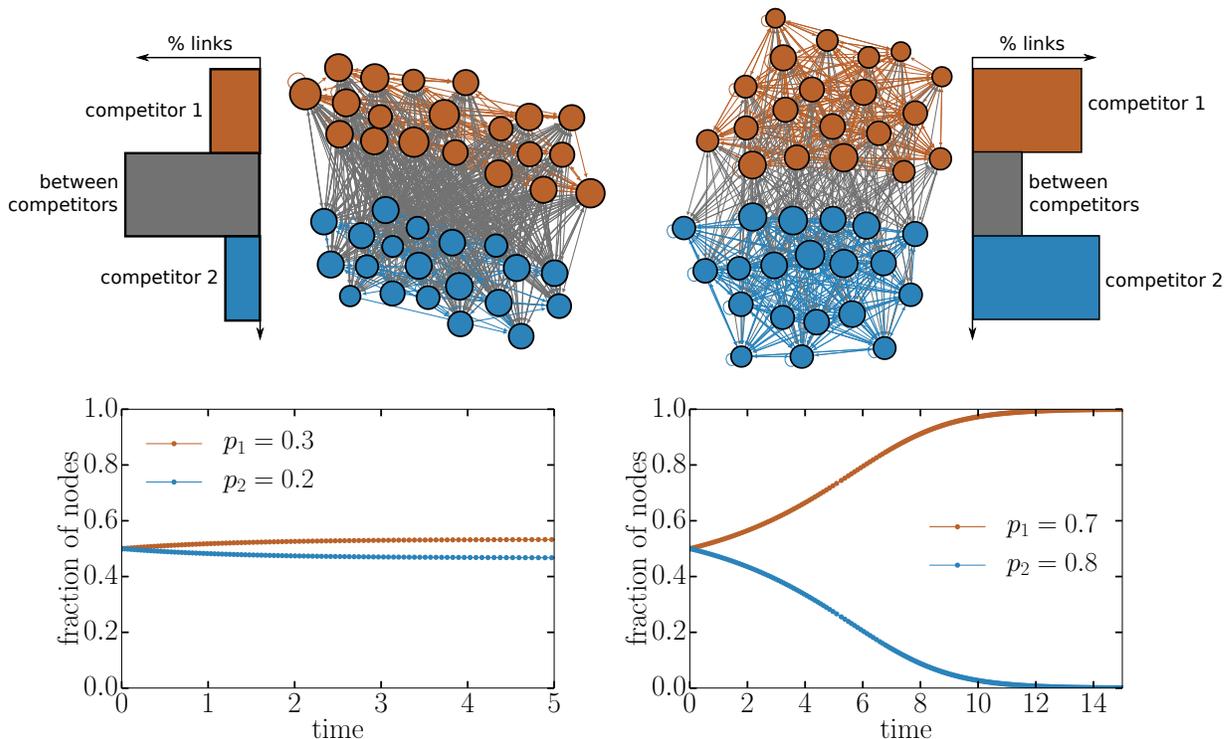

FIG. 2. **Examples of a two competitor contest with assortative and disassortative strategies.** We illustrate topologies given two set of strategies. Node color corresponds to its current state, node size is proportional to its total degree, and links are colored according to the states of the nodes they connect. (left) A disassortative structure: nodes tend to connect to nodes of a different state, notice that most links are grey which denotes inter block links. (right) An assortative structure: nodes tend to connect to nodes of the same state. (bottom) Time evolution of the voter model on these adaptive topologies. In the disassortative structure both competitors can coexist whereas the assortative structure leads to a winner-takes-all scenario.

The phenomenology of the prisoner's dilemma is thus very similar to that of our model: 2 fixed points corresponding to winner-takes-all scenarios, and a co-existence fixed point. In the general case without strategic trade-offs (see Supplementary Methods for a complete analysis), we can set $T = S = p_{12} = p_{21} \neq 0$ to force symmetry between the two strategies and avoid disconnected sub-populations. There is then a direct mapping between all fixed points when $\boldsymbol{A} = \boldsymbol{P}$.

This condition is not surprising since, unlike the prisoner's dilemma, the voter model does not allow one strategy to have an advantage over the other (i.e., a node converts its neighbour with a probability independent of their types; the outcome only depends on the relative number of neighbours of each type). The other key difference in the general forms of the two models is that ours includes network effects that co-evolve with the states of the nodes. A simple example from the mapping described above is when $p_{12} = p_{21} = 0$, where there are no dynamics whatsoever in the voter model since the two populations are disconnected; it is of course not the case for the well-mixed prisoner's dilemma even if $T = S = 0$.

There is a also a more subtle but fundamental distinction between the two models. In classic evolutionary game theory there are pairwise interactions between players where the payoff matrix determines who wins and thereby increases in relative frequency. The strategies adopted by players and the associated payoff matrices can only be inferred through population dynamics [21]. In contrast, in our model, the network structure (i.e., the directions and densities of the various edges) is shaped by and reflects the competing strategies at any point in time. As a consequence the competition and its dynamics can be inferred from a static snapshot of the network.

Despite these differences, there is striking similarity between the phenomenology observed in our model and in $g$-strategies game theory. This is likely due to the fact that the underlying dynamics in both models is determined by a term corresponding to the probability of interaction between two strategies and the associated reward/influence. In the following section we will consider the case of three competitors and their rich dynamical behaviours that can also be found in results of evolutionary game theory, but that emerge here for very different reasons. In fact, it will be much easier to interpret our results, and the tradeoffs from which they stem, in terms of the structure of the interaction network. In that sense, one significant advantage of our model is that the density matrix $\boldsymbol{P}$ is much less abstract than the payoff matrix $\boldsymbol{A}$ in the sense that it can be measured from relatively simple data. This network perspective thus allows us to apply our model to the previously discussed Twitter data (cf., end of the section and Fig. 8).

### The three competitor case

Having analyzed the case of two competitors, we now investigate the case of three competitors each trying to collect more nodes than the others. We assume for simplicity that competitors adopt the same strategy against all of their opponents. Thus, there is no distinction between opposition, only an "us and them" distinction. The constraints remain the same in that each competitor allocates a proportion $p_i$ of its strategy to reinforcing captured nodes, and the remaining $1 - p_i$ to pursuing nodes owned by its competitors. The elements $p_{ij}$ of the matrix $\boldsymbol{P}$ have the form

$$\boldsymbol{P} = \begin{pmatrix} p_1 & 1-p_1 & 1-p_1 \\ 1-p_2 & p_2 & 1-p_2 \\ 1-p_3 & 1-p_3 & p_3 \end{pmatrix}, \quad (7)$$

and the dynamics can be followed by equations of the form

$$\dot{x}_i = x_i \sum_{j \in \{1,2,3\}} \left[ \frac{p_{ij} x_j}{\sum_{l \in \{1,2,3\}} p_{lj} x_l} - \frac{p_{ji} x_j}{\sum_{l \in \{1,2,3\}} p_{li} x_l} \right], \quad (8)$$

for every $i \in \{1, 2, 3\}$, although one is superfluous as the system is constrained by $x_1 + x_2 + x_3 = 1$.

In the two-competitor scenario, the optimal strategy was $p = 1/2$. For example, if a competitor with this optimal strategy battled an opponent with $p < 1/2$, then the competition would finish with a mixed population where $p = 1/2$ held the majority. If, instead, the competitor with the optimal strategy battled two opponents each with a $p < 1/2$ then it would go extinct (see Fig. 3). Thus, the $p = 1/2$ strategy is not optimal in the three-competitor scenario.

To analyze the dynamics of the three-competitor contest, we note that setting any $x_i = 0$ constrains the phase space of the dynamics to the set of the remaining two competitors. Consequently, the results for the two-competitor contest apply directly to the three-competitor case. This yields six fixed points: three in which only one competitor exists and three mixed states with two competitors. There is another possible set of fixed points corresponding to coexistence of all three competitors, i.e., $x_1^*, x_2^*, x_3^* \neq 0$. We find that barring pathological cases, there can only exist at most one fixed point where all three competitors coexist (see Supplementary Methods). We can obtain this fixed point by removing the $x_i$ factor in all $\dot{x}_i = 0$ equations to eliminate solutions with any $x_i^* = 0$. Through a simple change of variables $z_i = x_i^* / \left( \sum_j p_{ji} x_j^* \right)$, the resulting system of equations can be written as the matrix equation

$$\boldsymbol{P}\vec{z} = \vec{1}, \quad (9)$$

where $\vec{z}$ is the vector of the new $z_i$ variables. The change of variables allows us to leverage the symmetries of the original equations. We can compute the remaining fixed point with three non-zero stable competitors by inverting $\boldsymbol{P}$ and solving a system of linear equations.

By computing the fixed points and determining their stabilities, we find that the three-competitor contest can be described by a set of five characteristic flow diagrams that we organize into four classes with qualitatively different behaviors and numbers/types of stable outcomes. Each class is distinguished by two simple features: (i) by the number of defensive strategies, i.e. how many competitors $i \in 1, 2, 3$ have $p_i \geq 1/2$; and (ii) by the number of pairwise combination of strategies that are generally defensive, i.e., whether $p_i + p_j$ is greater than 1 for pairings $\{i, j\}$ in $\{1, 2\}$, $\{1, 3\}$ and $\{2, 3\}$. Condition (ii) determines the dynamics along the edge of the $\vec{x}$ space (where $x_i = 0$ for exactly one $i$), and condition (i) informs us on the overall shape of the flow. In what follows, we discuss the dynamics of each class and its implications for three-competitor contests. We analyse the observed phenomenology in terms of the underlying network structure, but similar discussions exist for games with more than two strategies in evolutionary game theory [22, 23]

*Interior stable fixed point.* We already know that pairwise aggressive competitions promote coexistence in two-competitor cases. Similarly, with three competitors, all three strategies can coexist only when all strategies are offensive, i.e. $p_i < 1/2$ $\forall i$. This mixed fixed point is globally stable such that all trajectories lead to it provided that the frequencies of competitors are nonzero (see Fig. 4 for example). As a result, if two competitors have reached their equilibrium and a third competitor enters the contest at low abundance then the equilibrium is shifted to the interior fixed point. Thus, any third competitor with a strategy of $p < 1/2$ can successfully invade and reach a nonzero equilibrium frequency in the population. Following the results of Ref. [23] and knowing that the basic step of the voter model is a pairwise interaction (i.e., a two-player game), we know that there is only a single point of full co-existence.

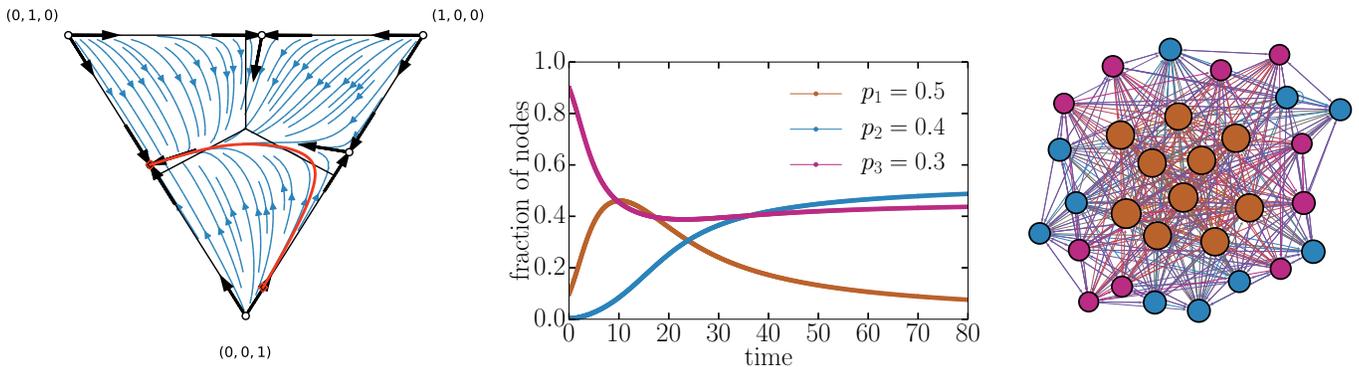

FIG. 3. **Dynamical trajectories for competitor using strategies $p_1 = 0.5$, $p_2 = 0.4$, and $p_3 = 0.3$.** (left) This triangle contains all possible dynamics: we show trajectories through a space where every point is defined by a unique $(x_1, x_2, x_3)$ state. Therefore, any point within the triangle correspond to a mixed state where all competitors have non-zero frequencies, whereas the edges correspond to two-competitor dynamics. All stable fixed points are shown as black dots, and semi-stable and unstable fixed points appear as open circles (their stability can also be deduced by the linear flows shown in black arrows around them). We delineate the three regions corresponding to states where one of the three competitors are respectively winning by a relative majority. While competitor 1 uses $p_1 = 0.5$, which is optimal on a one-on-one basis as shown by the two fixed points close to the $(1, 0, 0)$ apex, it systematically loses when all three strategies are involved. (middle) Example of a time series starting at $(0.099, 0.002, 0.899)$, which corresponds to the one highlighted in the left figure. (right) The network structure when two strategies are aggressive and one is defensive corresponds to a core-periphery structure with the core corresponding to the highest $p$ value. The network display style is the same as used in Fig. 2. The core is denser with nodes having a higher average total degree, but it does not target the periphery whereas nodes on the periphery preferentially targets the core and eventually win.

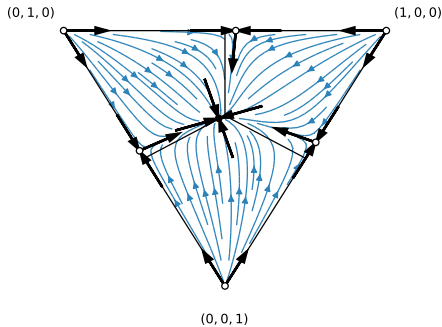

FIG. 4. **Coexistence of three aggressive strategies.** Depicted is the flow diagram of the voter dynamics given $p_1 = 0.3$, $p_2 = 0.2$ and $p_3 = 0.1$. As obtained through our analysis of the two competitor dynamics, aggressive strategies promote coexistence. This is generalized in the three competitor cases, where an interior stable fixed point can exist when $p_i < 1/2$ for all $i$.

*Single edge stable fixed point.* Another class of flow diagrams have a single final state in which two competitors coexist and one goes extinct. This occurs if only one strategy is defensive, i.e. $p_i \geq 1/2$ for only one $i$, which disrupts the fully mixed coexistence. In these cases, all trajectories with nonzero initial conditions lead to a fixed point corresponding to coexistence between the two offensive strategies. This class of flow diagrams has an interesting property that in some cases the winner of pairwise competitions is the loser in the three competitor contest (see Fig. 3 for an example of this behavior). This "winner turns loser" scenario is analogous to well-known paradoxes in voting systems with rational voters choosing between options [24]. To be clear, our results and methodology are distinct from the traditional framing of voter paradoxes. In such systems, there are rational voters choosing between options and voters have fixed preferences for the options. Various rules used to determine the winning option present paradoxical behavior. In our system, the analogous framing would be if nodes are voters and competitors are options. But since the networks we analyze are adaptive, voters are continually changing their preference for options. Despite the differences, the winner turns loser paradox emerges in our system through the existence of a stable fixed point between the pairwise "losing" strategies.

*Two stable fixed points.* There are two classes of competitions with two stable outcomes: either two stable mixed points where two strategies coexist, or one sta-ble mixed point and one stable winner-takes-all outcome. The former occurs when we have two defensive strategies but only one pairwise defensive combination. In this case, the two pairwise defensive match-ups will lead to two stable mixed points where the most aggressive strategy coexist with either of the defensive strategies. The second case with two stable points, which has one stable mixed state and one stable winner-takes-all outcome, occurs when there are two pairwise defensive match-ups. Obviously, the pairwise aggressive pair will still lead to a stable mixed state, but the two pairwise defensive pairs create unstable fixed points that delineate a basin of attraction for initial conditions leading to a winner-takes-all outcome for the most defensive strategy.

This class of flow diagrams contains trajectories in which a competitor approaches close to extinction before increasing to reach the majority of the population. These *long transients* appear to approach one stable set of fixed points before passing a critical threshold where the behavior rapidly changes (see Fig. 5). Interestingly, these trajectories also pass through areas in which one competitor is close to obtaining all of the nodes before eventually losing a majority of them. This type of paradoxical behaviour, where an initial decrease in the frequency of a competitor eventually leads to an enhancement of the same competitor, were originally observed in dynamics with cyclical dominance built-in the game structure (e.g. rock-paper-scissors) [25]. In our case it emerges because the winner of the pairwise match-ups can not simultaneously overcome two competitors with high frequencies, and must instead wait for one to be suppressed before taking over. A similar effect can also be caused by spatial constraints on predator-prey dynamics [26].

*Three stable fixed points.* Finally, when all pairings are defensive, then there are three stable fixed points corresponding to complete dominance by one competitor. In these competitions, there is no coexistence. The basins of attraction for the fixed points share borders such that small changes in initial conditions can completely change the outcome (see Fig. 6). The largest basin of attraction belongs to the competitor with the least defensive strategy, i.e., the lowest $p$. Note that only in contests between three defensive strategies, i.e. $p_i \geq 1/2 \; \forall i$, do we see an interior unstable fixed point as in Fig. 6.

**Generalized model and empirically-derived $P$ matrix**

Hithereto, we primarily considered particular strategic tradeoffs between offense and defense, but our framework is much more general. Competitors may distinguish between different opponents and split their offensive efforts. For instance, consider the following strategy matrix

$$\boldsymbol{P} = \begin{pmatrix} 0.33 & 0.32 & 0.35 \\ 0.35 & 0.33 & 0.32 \\ 0.32 & 0.35 & 0.33 \end{pmatrix} , \qquad (10)$$

in which we have built a cyclical structure. Competitor 1 targets competitor 3 preferentially, competitor 2 targets competitor 1, and competitor 3 targets competitor 2. Even though the preferential targeting is small, there is enough asymmetry to push the system towards cyclical Rock-Paper-Scissor dynamics as shown in Fig. 7. This kind of behavior has been well-studied [27–30] and is easily generated when competitors distinguish between opposition. Although the trajectories that lead to the final steady state can be very different, the final solutions derived in our analysis (see Supplementary Methods) still holds and can apply to an arbitrary number of competitors.

To illustrate how one might apply our general framework to real world scenarios, we revisit the empirical data from Twitter on political discussions presented in Fig. 1. The echo chambers observed in the political discussions imply that users are following defensive strategies, in which they devote more of their effort to like-minded users than users with opposing views. The shape of the data suggests a parametrization in terms of three ideologies: liberal, centrist, and conservative. Coarse-graining the empirical matrix in terms of this parametrization directly gives a $\boldsymbol{P}$ matrix for its competitor dynamics. Figure 8 summarizes the prediction of our model based on this empirical $\boldsymbol{P}$ matrix. As expected from the defensive strategies and our previous analysis, the dynamical system finds itself in a regime sensitive to initial conditions with possibly long transient behavior. Uncertainty in initial conditions is thus not only reflected in terms of which competitor ultimately wins, but also potentially in how long it will take before a winner emerges.

Figure 8 shows how our model incorporates empirical data when each competitor can distinguish between opposing competitors. In such cases, the extinction of one competitor does not reduce the competition to an equivalent two-competitor scenario. Indeed, in Figure 8 after the centrist ideology falls to zero, conservatives have higher outgoing degree than liberals as liberals waste edges targeting non-existent centrists. Re-scaling strategies to account for the disappearance of a competitor can lead to a completely different outcome, see dotted curves on Fig. 8. Assuming the competitors are informed of the current state of the system, this opens the door for more complex strategies that would themselves adaptively co-evolve with the population.

## DISCUSSION

There is a rich history in both biology and social sciences of mathematical models used to understand the dynamics of competition over finite resources [1–8]. The canonical class of these models, which include voter models and Moran processes, do not incorporate characteristic features of many real-world competitions. Motivated by empirical data from Twitter and other online forums, we extend these models by adding three features: 1. com-petitors can adopt different strategies towards resources they control and those they do not, 2. there is a tradeoff between offense (seeking resources to acquire) and defense (protecting resources in possession), and 3. the competition structure adapts according to the interplay between strategies.

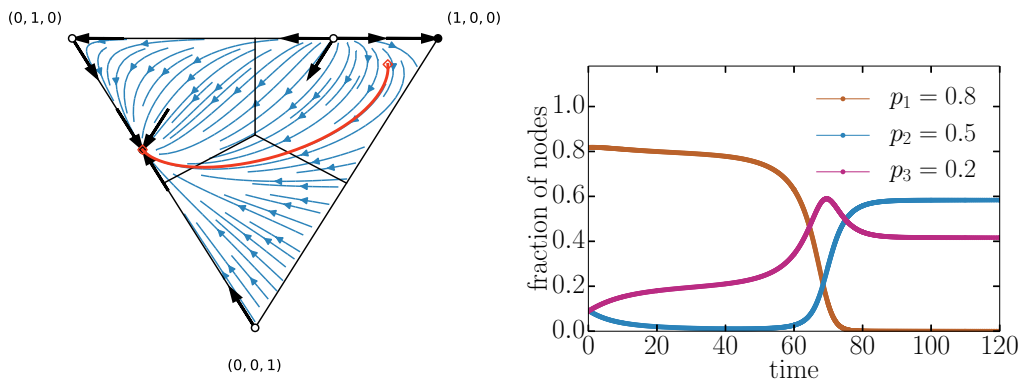

FIG. 5. **Long transient behavior with mixed strategies.** (left) We compute the flow diagram using $p_1 = 0.8$, $p_2 = 0.5$ and $p_3 = 0.2$. Notice that the pairwise competition of competitor 1 and 3 leads to a line of fixed points on the right edge of the simplex because of the pathological case $p_1 = 1 - p_3$, see Supplementary Methods for details. (right) Example of a time series where the final winner ($p_2$) stays close to extinction until competitor 3 obtains a majority.

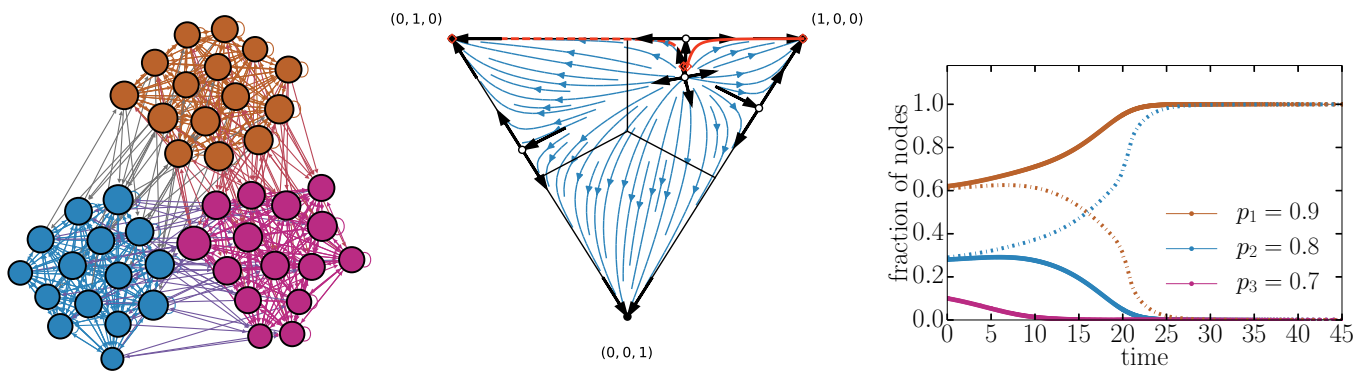

FIG. 6. **Dynamical trajectories in a modular competition between $p_1 = 0.9$, $p_2 = 0.8$, and $p_3 = 0.7$.** (left) Example of a network built from three defensive strategies leads to three distinct modules. The network display style is the same as used in Fig. 2. (middle) Flow diagram of the voter model dynamics given these three defensive strategies shows the three basins of attractions. (right) Example of two time series with slightly different initial conditions shows how the final outcomes of the competition can change. The one with full markers starts at $(0.62, 0.28, 0.1)$ and the one shown with a dotted line starts at $(0.61, 0.29, 0.1)$.

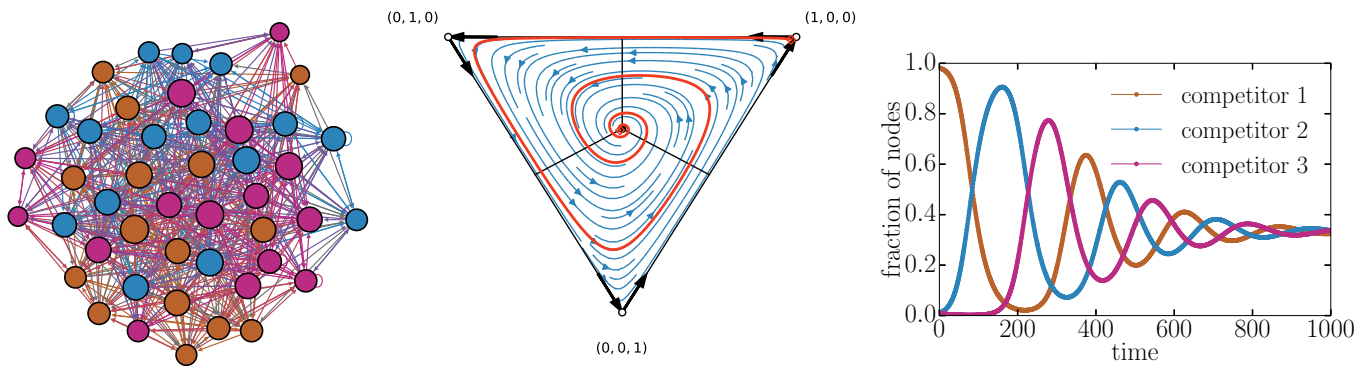

FIG. 7. **Rock-paper-scissor behavior in a well-mixed system with slight asymmetries.** (left) The network architecture is well-mixed (i.e. an homogeneous network) as the asymmetries in density between groups are of the order of 1 in a 100 links. The network display style is the same as used in Fig. 2. (middle) The flow diagram resulting from the $\boldsymbol{P}$ matrix in Eq. (10) shows cyclic behavior. (right) The slight cyclical structure (1 targets 3, 2 targets 1 and 3 targets 2) is enough to give rise to damped cyclical behavior. Note that an asymmetry in the initial conditions is also needed. The stronger the asymmetries, the longer the oscillatory transient would be.

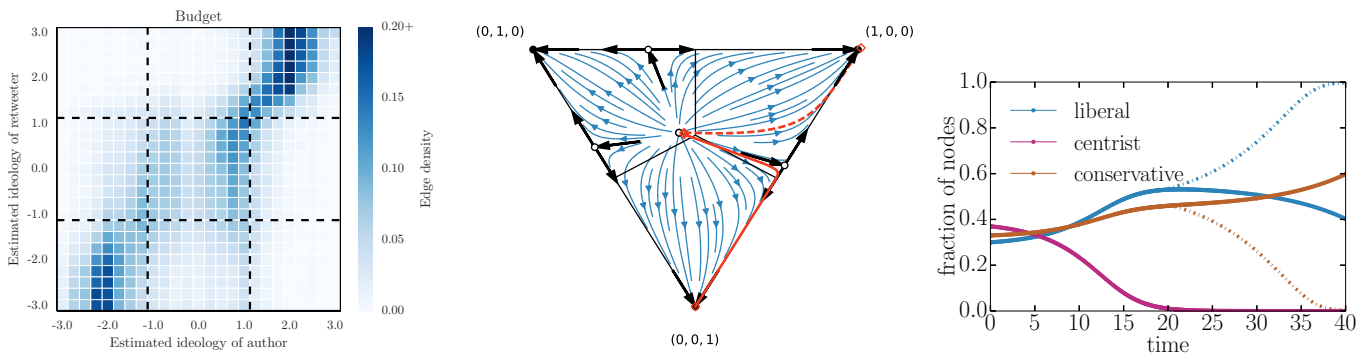

FIG. 8. **Application of our model to Twitter data.** (left) Shown are strategies of retweets for different inferred ideologies of authors, according to (and normalized) for every inferred ideologies of retweeters using the federal budget data. The structure of strategies leads to a natural parametrization in terms of three ideologies showed by dotted lines: liberal (smaller than -1.0), centrist (between -1.0 and 1.0) and conservative (above 1.0). Coarse-graining of retweets over this 3 by 3 matrix, then normalizing per row, leads directly to a possible $\boldsymbol{P}$ matrix. (middle) We obtain $\boldsymbol{P}$ matrix and compute the flow diagram with our general analysis (see Supplementary Methods). We highlight two possible time series to illustrate how small changes in initial conditions can lead to drastically different outcomes. (right) One possible time series (highlighted in solid line on the flow diagram) shows the transient behavior. The dotted curves do not exactly correspond to the dotted curve in the flow diagram, but rather to a scenario where strategies are re-scaled (or re-normalized) to remove the centrist ideology from the system once its frequency falls below 1% of the population.

In competitions between two competitors there is a single optimal strategy that invests equally in defense and offense. The addition of another competitor creates a much richer set of dynamics with four qualitative regimes. One immediate consequence is the absence of an optimal strategy. The different regimes in the three-competitor case exhibit behavior observed in real-world competitions, including winner-turn-loser paradoxes and instances where the ultimate winning competitors must first past close to extinction.

When there are three competitors, coexistence can only occur if all competitors adopt primarily offensive strategies. When competitors adopt defensive strategies, they promote winner-takes-all outcomes. This outcome is surprising considering the network structure that results from such competitions. Defensive strategies lead to high modularity in which each competitor acquires a set of resources/nodes and forms many links between the nodes in order to maintain possession. This network structure seems like it would lead to coexistence because each competitor protects a set of nodes and makes little effort to acquire others. However, it is precisely this structure that leads to the extinction of one or two competitors. The main reason is that there is a strong positive feedback loop, where once there is a disparity in the number of nodes belonging to each competitor, the majority owner is more likely to acquire new nodes even if it spends little effort trying to do so. In contrast, a competition with only offensive strategies produces a network structure in which nodes are continually changing hands. Although this structure seems unstable, the average number of nodes possessed by competitors reaches a non-zero steady state, i.e. coexistence.

If we consider the competitor who claims the most nodes/resources to be the winner, then we find that the optimal strategy in the two-competitor case can easily be out-competed in the three-competitor case. Namely, the $p = 1/2$ strategy beats an either more offensive or defensive strategy, but loses when facing two offensive or defensive competitors. Instead the strategy with a value of $\boldsymbol{P}$ in between the others tends to win. The best strategy, therefore, is to be the second most aggressive competitor. Thus, the best strategy for a given competitor depends on the strategies adopted by the other competitors.

The dependence of the winning strategy on the competing strategies gives rise to well-known voting paradoxes. For example Fig. 3 illustrates the *Condorcet Winner* and the *Violation of the subset choice condition* paradoxes (CW and SCC), which are some of the most frequently occurring paradoxes [24]. The CW paradox occurs when a competitor loses an election despite the fact that it would be preferred over any of the competing alternatives. In our model, this occurs in competitions between offensive strategies. On a one-on-one basis the strategy closest to $p = 1/2$ would win, but it is the middle strategy that wins in three-competitor scenarios. So, for example, in a competition between $p_1 = 1/2, p_2 = 1/3$, and $p_3 = 1/4$, the $p_1$ strategy would win any pairwise competition but the $p_2$ strategy would win in the three competitor case. This dynamic also implies the SCC paradox in which the expected winner of an election may eventually lose following the removal of a current loser. Indeed, the second most offensive strategy is expected to win if all competitors are present, but loses if the most offensive strategy is removed. The inverse of this behavior which also exists in our model system is the decoy or asymmetric dominance effect [31] whereby introducing an inferior option/competitor can artificially promote another option/competitor, even if it was not an initial favorite. Interestingly, these paradoxes were primarily identified in social choice systems with very different features than exist in our model. Yet, we find similar paradoxes are produced by the interplay between strategic tradeoffs and adaptive network structure, without the need for any active choice.

Another interesting feature found in our model is the existence of extremely long transient dynamics. An example displayed in Fig. 5 shows that an eventual winner passes very close to zero frequency in the population corresponding to extinction or complete loss. Interestingly, the competitor who was leading initially ends up going extinct. If mid-competition we were to remove the eventual winner, then the initially leading competitor would win. These long dynamics have important implications when we consider political campaigns. A political campaign that steadily decreased in the polls until it claimed only a small percentage of support would generally be assumed to be a lost cause. Yet in our simple model, such a campaign may simply be following a long transient and may ultimately win despite its seeming initial failure. We note that strategies in our model are fixed so that if a campaign with the long transient wins, it is not due to any strategic shift or adjustment.

We also note that the competitions analyzed in our paper are not just decided by strategy but also by initial conditions. In fact, the expected dynamics and eventual winner can be extremely sensitive to initial conditions. For example, if all competitors adopt defensive strategies then there are three stable fixed points corresponding to the three different winner-takes-all outcomes. Each stable fixed point has its own basin of attraction separated by unstable fixed points. One of these unstable fixed points corresponds to an initial condition where all three competitors are initially present. As observed in Fig. 6, small deviations around this point push the dynamics to any of the three winner-takes-all outcome.

In this paper, we highlighted many similarities between our results and evolutionary game theory. These likely stem from the fact that in both cases strategies interact according to probabilities and there is a resultant payout or reward. While the underlying dynamics share similar elements, there are larger conceptual differences. Our model features an adaptive network structure which means that, at any time, the network encodes the strategies. One can look at a node and observe the outgoing and ingoing edges to infer the $P$ matrix and compute the subsequent dynamics—as we did in our Twitter analysis. In contrast, classic evolutionary game theory often does not include any population structure apart from relative frequencies of players (and those that do often consider fixed structures). As a result, in order to determine the competition dynamics one must know the payoff matrix and the probabilities that certain strategies are played. This data is typically hidden from view in empirical systems and challenging to infer without population dynamics. Thus, despite the qualitative connection observed between the payoff structure of game theory and the connectivity structure of our adaptive networks, empirical analysis is likely more readily accessible using our conceptualization and approach.

Finally, our analysis focused on a particular type of tradeoff but it is certainly not the only one. For example, the Twitter data we used to fit a $P$ matrix did not follow that particular tradeoff. Nonetheless, our general analysis still holds and we found similar qualitative dynamics as observed in our more restricted three competitor case. There are, of course, many caveats involved in the use of online discussion forum data, and it might be that incorporating data from other sources such as Facebook or Reddit might yield different $P$ matrix structures. Although individuals have access to many online platforms and activities, their behaviors are likely correlated with their ideology. Consequently, interpreting the data from online discussion platforms is an active area of research [11, 12, 32]. Our mathematical model complements this work by providing a simple parametrization concerning offense/defense strategies that can be tuned to multiple data sources. Not only does it exhibit behavior found in complex real world scenarios but there are many interesting open questions. For example, do defensive strategies really lead to winner-takes-all competitions while offensive strategies promote coexistence of competitors/ideas? Do the qualitative dynamics change if competitors face different constraints? What happens if competitors alter their strategies over the course of the competition? At the very least, our analytical results provide examples of what to look for in further study.

### ACKNOWLEDGEMENTS

We thank Marion Dumas, Mirta Galesic, Sidney Redner and anonymous reviewers for their comments, as well as Pablo Barberá for discussing and sharing the Twitter results of Figs. 1 and 8. This work has been supported by the Santa Fe Institute, the James S. McDonnell Foundation Postdoctoral Fellowship (L.H.-D.), the National Science Foundation under Grant DMS-1622390 (L.H.-D.), the Santa Fe Institute Omidyar Postdoctoral Fellowship (E.L.), the Fonds de recherche du Québec–Nature et technologies (J.-G.Y., A.A.) and the Defense Threat Reduction Agency Basic Research Award HDTRA1-10-1-0088 (P.-A.N.).

### AUTHOR CONTRIBUTIONS

L.H.-D., A.A., P.-A.N., J.-G.Y., and E.L. conceived of the model, performed analysis and interpretation of results, and wrote the manuscripts.

## COMPETING FINANCIAL INTERESTS

The authors declare that they have no competing financial interests.

---

# Strategic tradeoffs in competitor dynamics on adaptive networks [Supplementary Methods]

Laurent Hébert-Dufresne,[1, 2] Antoine Allard,[3] Pierre-André Noël,[4] Jean-Gabriel Young,[5] and Eric Libby[1, *]

[1]*Santa Fe Institute, Santa Fe, NM 87501, USA*
[2]*Institute for Disease Modeling, Bellevue, WA, 98005, USA*
[3]*Centre de Recerca Matemàtica, E-08193 Bellaterra (Barcelona), Spain*
[4]*University of California, Davis CA 95616, USA*
[5]*Département de physique, de génie physique et d'optique,
Université Laval, Québec (Qc), G1V 0A6, Canada*

This document provides a full analysis of the model described in the main text. We provide general mean-field equations for the dynamics in which an arbitrary number of competitors are at play. We describe all possible outcomes of the case with two competitors, and solve all fixed points of the dynamics in the case with an arbitrary number of competitors. We also reproduce the full data from Ref. [13] in main text, use them to estimate our model parameters and produce the corresponding flow diagrams.

## I. GENERAL FORMULATION OF THE MODEL

We consider a population of $N$ nodes belonging at any given time to one of a set of different competitors, noted $\mathcal{G}$. At time $t$, $N_i(t)$ nodes belong to competitor $i \in \mathcal{G}$ during the interval $[t, t+dt]$. The interactions between the nodes are prescribed by the *directed* stochastic block model specified by the matrix $\boldsymbol{P}$ whose elements $p_{ij}$ correspond to the probability that a directed link exists from a node of state $i$ towards a node of state $j$. Nodes continuously change hands by randomly switching to the state of one of the nodes that have a directed link pointing to them (i.e., incoming link). Hence, at any time $t$, a node of state $i$ has $k_{ij}^{\text{in}}(t)$ incoming links from nodes of state $j$ with probability

$$\Pr[k_{ij}^{\text{in}}(t) = n] = \binom{N_j(t)}{n} p_{ji}^n (1-p_{ji})^{N_j(t)-n} \; , \qquad (1)$$

and opts for state $j$ with a probability $k_{ij}^{\text{in}}(t) / \sum_{l \in \mathcal{G}} k_{il}^{\text{in}}(t)$. Once a node has changed state, its incoming and outgoing links are redrawn according to the matrix $\boldsymbol{P}$. Note that nodes may stick to their current state since self-loops are allowed (i.e., stubbornness) and incoming links may be from nodes sharing the same state (i.e., group cohesion).

Let us now state the general mean-field equations to follow the dynamics in the limit of an infinite population being disputed by a set $\mathcal{G}$ of different competitors. To do so, we consider an *annealed* version of the model in which links are continuously redrawn according to matrix $\boldsymbol{P}$ regardless of whether nodes change their state or not, and this is achieved at a rate faster than the process of switching states itself (i.e., there is no temporal correlation). We also consider the limit $N \to \infty$ and define $x_i(t) = N_i(t)/N$ as the fraction of nodes sharing state $i$. Note that we will explicitly write the time dependency only if required to avoid confusion. A node of state $i$ has a number of incoming links from nodes of state $j$ proportional to $p_{ji} x_j$, and switched its state to $j$ at a rate proportional to $p_{ji} x_j / \sum_{l \in \mathcal{G}} p_{li} x_l$. Altogether, the time evolution of the system is described by the following differential equation

$$\dot{x}_i = x_i \sum_j x_j \bigl( p_{ij} \varphi_j - p_{ji} \varphi_i \bigr) \qquad \forall \; i \in \mathcal{G} \qquad (2)$$

where the elements of the matrix $\boldsymbol{P}$, $\{p_{ij}\}_{i,j \in \mathcal{G}}$, are the probability that a directed link exists from a node of competitor $j$ towards a node of competitor $i$, and where $\varphi_j$ is the reciprocal of the expected in-degree of nodes of competitor $j$

$$\varphi_j = \frac{1}{\sum_l p_{lj} x_l} \; . \qquad (3)$$

---

* Correspondence to : elibby@santafe.edu

We can readily compute the elements of the Jacobian matrix from which the stability of the fixed points of Eq. (2) can be determined

$$J_{ki} = \frac{\partial \dot{x}_k}{\partial x_i} = x_k \sum_j \frac{\partial x_j}{\partial x_i}(p_{kj}\varphi_j - p_{jk}\varphi_k) + \frac{\partial x_k}{\partial x_i}\sum_j x_j(p_{kj}\varphi_j - p_{jk}\varphi_k) + x_k \sum_j x_j \left(p_{kj}\frac{\partial \varphi_j}{\partial x_i} - p_{jk}\frac{\partial \varphi_k}{\partial x_i}\right)$$

$$= x_k(p_{ki}\varphi_i - p_{ik}\varphi_k) + \delta_{ik}\sum_j x_j(p_{kj}\varphi_j - p_{jk}\varphi_k) + x_k \sum_j x_j(p_{jk}p_{ik}\varphi_k^2 - p_{kj}p_{ij}\varphi_j^2) \ . \quad (4)$$

Let us finally rewrite Eq. (2) in a way that will be useful in the subsequent analysis.

$$\dot{x}_k = x_k\left(\sum_j p_{kj}x_j\varphi_j - 1\right) \ . \quad (5)$$

## II. GENERAL SOLUTION FOR THE CASE $|\mathcal{G}| = 2$

Let us consider the two-competitor scenario, $\mathcal{G} = \{1,2\}$, first without any constraint on the elements of the matrix $\boldsymbol{P}$. The conservation of nodes, $x_1 + x_2 = 1$, implies that the dynamics of the system, Eq. (2), can be tracked with the single equation

$$\dot{x}_1 = x_1(1-x_1)\left[\frac{1}{(1-\alpha_2)x_1 + \alpha_2} - \frac{1}{(\alpha_1-1)x_1 + 1}\right] \ , \quad (6)$$

where we have defined $\alpha_1 = p_{11}/p_{21}$ and $\alpha_2 = p_{22}/p_{12}$. Note that $\alpha_1, \alpha_2 \geq 0$ since the elements of matrix $\boldsymbol{P}$ are probabilities. Note also that the system is static regardless of its initial conditions whenever $\alpha_1 = \alpha_2 = 1$. From Eq. (6), we can readily identify the three possible fixed points

$$x_1^{*(1)} = 0 \ ; \qquad x_1^{*(2)} = 1 \ ; \qquad x_1^{*(3)} = \frac{1}{1 + \frac{(\alpha_1-1)}{(\alpha_2-1)}} \ , \quad (7)$$

with $0 \leq x_1^{*(3)} \leq 1$ whenever

$$\frac{(\alpha_1-1)}{(\alpha_2-1)} \geq 0 \ . \quad (8)$$

From Eq. (4), we see that the elements of the Jacobian matrix become

$$J_{11} = x_2(\beta_{12} - \beta_{21}) + x_1 x_2(\beta_{11}\beta_{21} - \beta_{12}^2) \quad (9a)$$
$$J_{12} = x_1(\beta_{12} - \beta_{21}) - x_1 x_2(\beta_{12}\beta_{22} - \beta_{21}^2) \quad (9b)$$
$$J_{21} = -x_2(\beta_{12} - \beta_{21}) - x_1 x_2(\beta_{11}\beta_{21} - \beta_{12}^2) \quad (9c)$$
$$J_{22} = -x_1(\beta_{12} - \beta_{21}) + x_1 x_2(\beta_{12}\beta_{22} - \beta_{21}^2) \quad (9d)$$

where we have defined $\beta_{ij} = p_{ij}\varphi_j$. Noting that $J_{11}J_{22} - J_{12}J_{21} = 0$, we find that the eigenvalues, $\lambda$, of the Jacobian matrix are the solution of

$$\lambda(J_{11} + J_{22} - \lambda) = 0 \ . \quad (10)$$

The first eigenvalue $\lambda_1 = 0$ is the result of the conservation of the nodes limiting the dynamics to a single line in the phase space $(x_1, x_2)$. The second eigenvalue, $\lambda_2$, is

$$\lambda_2 = (x_2 - x_1)(\beta_{12} - \beta_{21}) + x_1 x_2(\beta_{11}\beta_{21} + \beta_{12}\beta_{22} - \beta_{12}^2 - \beta_{21}^2) \ . \quad (11)$$

Substituting Eq. (7) into this last result, we find that each fixed point is stable whenever

$$\lambda_2^{(1)} = \frac{1}{\alpha_2} - 1 < 0 \quad (12a)$$

$$\lambda_2^{(2)} = \frac{1}{\alpha_1} - 1 < 0 \quad (12b)$$

$$\lambda_2^{(3)} = x_1^{*(3)}\left(1 - x_1^{*(3)}\right)\beta_{12}^2(\alpha_1 + \alpha_2 - 2) < 0 \ . \quad (12c)$$

Using these results, the condition (8) can be written as

$$\frac{\alpha_1 \lambda_2^{(2)}}{\alpha_2 \lambda_2^{(1)}} \geq 0 \tag{13}$$

which implies that $x_1^{*(1)}$ and $x_1^{*(2)}$ have the same stability whenever $0 \leq x_1^{*(3)} \leq 1$ and have different stabilities otherwise. From Eq. (12c), we see that $x_1^{*(3)}$ is stable when

$$\alpha_1 \lambda_2^{(1)} + \alpha_2 \lambda_2^{(2)} > 0 , \tag{14}$$

which in turn implies that $x_1^{*(3)}$ has a different stability than $x_1^{*(1)}$ and $x_1^{*(2)}$ whenever $x_1^{*(3)} \in [0, 1]$.

### III. SOLUTION FOR $|\mathcal{G}| = 2$ WITH STRATEGIC TRADEOFF

To emulate limited resources, we constrain the matrix $\boldsymbol{p}$ to the following structure

$$\boldsymbol{p} = \begin{pmatrix} p_1 & 1 - p_1 \\ 1 - p_2 & p_2 \end{pmatrix} , \tag{15}$$

where $0 \leq p_1, p_2 \leq 1$ are two free parameters. Combining Eq. (15) and the conservation condition $x_2 = 1 - x_1$, we obtain that the differential equation describing the dynamics of the $|\mathcal{G}| = 2$ scenario is

$$\dot{x}_1 = \frac{x_1(1 - x_1)(1 - p_1)}{x_1(1 - p_1) + (1 - x_1)p_2} - \frac{x_1(1 - x_1)(1 - p_2)}{x_1 p_1 + (1 - x_1)(1 - p_2)} , \tag{16}$$

From the complete analysis of this model given in Sec. II, we readily identify the three fixed points

$$x_1^{*(1)} = 0 , \tag{17a}$$

$$x_1^{*(2)} = 1 , \tag{17b}$$

$$x_1^{*(3)} = \left(1 + \frac{1 - p_1}{1 - p_2}\right)^{-1} . \tag{17c}$$

and draw the following conclusions:

- If $p_1 + p_2 < 1$, both $x_1^{*(1)}$ and $x_1^{*(2)}$ are unstable and $x_1^{*(3)}$ is stable:
  - Competitor 1 wins majority if $p_1 > p_2$;
  - It is a tie if $p_1 = p_2$;
  - Competitor 2 wins majority if $p_1 < p_2$.

- If $p_1 + p_2 = 1$, the system is static—$x_1(t) = x_1(0)$ for all time $t$—and the winning strategy is trivially determined by the initial conditions:
  - Competitor 1 wins majority if $x_1(0) > 1/2$;
  - It is a draw if $x_1(0) = 1/2$;
  - Competitor 2 wins majority if $x_1(0) < 1/2$.

- If $p_1 + p_2 > 1$, both $x_1^{*(1)}$ and $x_1^{*(2)}$ are stable and $x_1^{*(3)}$ is unstable:
  - Competitor 1 wins by unanimity if $x_1(0) > \left(1 + \frac{1-p_1}{1-p_2}\right)^{-1}$;
  - No draw is possible;
  - Competitor 2 wins by unanimity if $x_1(0) < \left(1 + \frac{1-p_1}{1-p_2}\right)^{-1}$.

# IV. ANALYSIS OF FIXED POINTS IN THE CASE $|\mathcal{G}| > 2$

We change our focus to a scenario where more than two competitors compete to rally a majority of nodes. Eqs. (5) are non-linear, and their fixed points could be non-trivial in principle. However, from Eq. (2) we see that setting $x_i = 0$ constrains the phase space of the dynamics to the set of the remaining $\mathcal{G}\backslash i$ competitors. In other words, all pairwise fixed points (i.e., fixed points with at most two non-zero $x_k^*$) are already known. With $|\mathcal{G}| = 3$, the missing fixed points thus correspond to steady states where $x_k^* > 0 \ \forall \ k \in 1, 2, 3$; and so on as we add competitors. It turns out that (a) we can find the *unique* bulk fixed point of the $|\mathcal{G}|$ competitors case and (b) build the other fixed points recursively from the $|\mathcal{G}| - 1$ competitors cases, all the way down to $|\mathcal{G}| = 2$. For the sake of clarity, we define $g = |\mathcal{G}|$ in what follows.

## A. Bulk fixed point

We define *bulk* fixed points as vectors $\vec{x}^*$ whose elements $x_k^* > 0 \ \forall \ k$, and for which $\dot{x}_k\big|_{\vec{x}^*} = 0 \ \forall \ k$. In this section, we obtain an analytical expression for the bulk fixed points and show that only a *single* one exists for each $g = 2, 3, ....$

By definition of the bulk fixed point, one must exclude $x_k^* = 0$ from $\vec{x}^*$, for all $k$. This is done by factoring out the leading $x_k$ (i.e., the $x_k^* = 0$ root) from Eq. (5). We are left with the simpler system

$$\dot{x}_k = \sum_j p_{kj} x_j^* \varphi_j^* - 1 = 0 \qquad k = 1, 2, .., g \ . \tag{18}$$

Notice that Eq. (18) can be written in matrix form as

$$\boldsymbol{P} \boldsymbol{z} = \boldsymbol{1} \tag{19}$$

where $\boldsymbol{P}$ is the $g \times g$ matrix of parameters $\{p_{ij}\}$ and where the $k^{\text{th}}$ element of $\boldsymbol{z}$ is defined as $z_k := x_k^* \phi_k^*$. Provided that $\boldsymbol{P}$ is invertible, one finds

$$\boldsymbol{z} = \boldsymbol{P}^{-1} \boldsymbol{1} \quad \iff \quad x_k^* \phi_k^* = s_k \qquad k = 1, 2, .., g \ , \tag{20}$$

where $s_k$ is defined as the sum of the $k^{\text{th}}$ row of the invert of $\boldsymbol{P}$. It then becomes apparent that we have removed the non-linear terms of Eq. (2); multiplying by $(\varphi_k^*)^{-1}$ on both sides yields a system of linear equations for $\boldsymbol{x}^*$:

$$x_k^* - s_k \left( \sum_j p_{jk} x_j^* \right) = 0 \qquad k = 1, 2, ..., g \ . \tag{21}$$

Note that in solving (21), we have no guarantee that $\sum_j x_j^* = 1$. It could therefore be that $\boldsymbol{x}^*$ is not a distribution. This problem is addressed by removing a redundant equation for $x_g^*$ (without loss of generality), and using the normalization condition

$$\sum_{j=1}^{g} x_j^* = 1 \quad \iff \quad x_g^* \equiv 1 - \sum_{j<g} x_j^* \ . \tag{22}$$

We are then guaranteed that the solutions of the resulting equation,

$$x_k^* - s_k \sum_{j<g} (p_{jk} - p_{gk}) x_j^* = s_k p_{gk} \qquad k = 1, 2, ..., g - 1 \ , \tag{23}$$

are *normalized* bulk fixed point. Eq. (22) can be used to recover $x_g^*$ explicitly. The solution is unique as long as the system (23) is independent. The solution *can* be a non-bulk fixed point if the point is degenerated. Without loss of generality, suppose $x_{g-1}^* = 0$, such that the $g$-competitors "bulk-point" actually is not non-zero for all $k$. Equation (23) tells us that this can only happen if

$$s_{g-1} p_{g,g-1} = 0 \ , \tag{24}$$

i.e. if the sum of the $(g-1)^{\text{th}}$ row of the $\boldsymbol{P}^{-1}$ is zero. The other possibility, $p_{g,g-1} = 0$, can be avoided by removing a different row in Eq. 22, and will only lead to unavoidable degenerated bulk point in the $\ell^{\text{th}}$ dimension if $p_{j,\ell} = 0 \ \forall \ j$. Note that solutions *can* still reside outside of the simplex, i.e. $x_j^* < 0$ for some $j$, which we ignore as they are non-physical fixed points.

### B. On fixed points recurrence

Let us consider, without loss of generality, that $x_g^* = 0$ and $\sum_{j<g} x_j^* = 1$. Then Eqs. (18) become

$$\dot{x}_k\bigg|_{\vec{x}^*} = x_k^* \left[\sum_{j<g} x_j^* p_{kj} \varphi_j - 1\right] \qquad k = 1, 2, .., g-1 ,\tag{25a}$$

$$\dot{x}_g\bigg|_{\vec{x}^*} = 0 ,\tag{25b}$$

$$\varphi_j\bigg|_{\vec{x}^*} := \left[\sum_{\ell<g} x_\ell^* p_{\ell j}\right]^{-1} .\tag{25c}$$

where $\vec{x}^*$ is here any fixed point which satisfies the criterion stated above. We immediately recover Eqs. (18) for $g-1$ competitors by removing the trivial Eq. (25b), and redefining the summation limits. Therefore, given a $g$ competitors system, one can obtain all fixed points where one of the components is null from the $g-1$ competitors system, recursively, all the way down to $g = 2$ which we fully solved previously.

To enumerate all fixed points, we therefore only need to

1. Find the bulk point for $g$ competitors.

2. For all $\ell = 1, 2, ..., g$, set $x_\ell^* = 0$ and go back to step 1 with $g' = g - 1$.

The procedures stops once we reach $g' = 2$.

### C. Scenario with strategic tradeoffs

In the scenario with strategic tradeoff, we have

$$p_{rs} = p_r \delta_{rs} + (1 - p_r)\bar{\delta}_{rs} \equiv (1 - p_r) + \delta_{rs}(2p_r - 1) .\tag{26}$$

The resulting $\boldsymbol{P}$ matrices are full rank matrices whenever there is at most one $p_r = 1/2$; all rows are independent from one another because of their diagonal element. Incidentally, any pair of competitors with both $p_r = 1/2$ will remain at their initial conditions as in the case with two competitors. In that case, we do not expect a single bulk point, but a *hyper-plane* of fixed points that connects $g - 2$ edges parallel to the competitors pair edge, along every faces of the simplex (see Fig. 1 for example).

For all other cases, the inverse of $\boldsymbol{P}$ can be calculated exactly using the Sherman-Morrison formula and the matrix representation $\boldsymbol{P} = [2\,\mathrm{diag}(\boldsymbol{P}) - \boldsymbol{I}] + (\boldsymbol{1} + \boldsymbol{P})\boldsymbol{1}^T$, where $\boldsymbol{P}$ is the column matrix of probabilities $p_r$ and $\boldsymbol{1}$ is a column matrix of 1. One finds

$$[\boldsymbol{P}^{-1}]_{rs} = \frac{\delta_{rs}}{2p_r - 1} - \frac{(1 - p_r)}{(2p_r - 1)(2p_s - 1)}\left[1 + \sum_j \frac{1 - p_j}{2p_j - 1}\right]^{-1}\tag{27}$$

Combining the recursive strategy of § IV B and Eq. (23) yields the position of all fixed points. Moreover, Eq. (27) allows us to compute $s_k$—we simply sum over each row of the inverse matrix—and obtain a condition for $s_k = 0$, i.e., a condition that predicts whether the "fixed bulk point" is degenerated and actually lives on the edges of the $g$-simplex (see Eq. 24). Doing so, we obtain the following condition for the non-existence of bulk point in the $g$ competitors case (with strategic tradeoffs)

$$\sum_j \frac{p_k}{2p_j - 1} = \sum_j \frac{p_j}{2p_j - 1} \qquad \text{for any } k = 1, .., g.\tag{28}$$

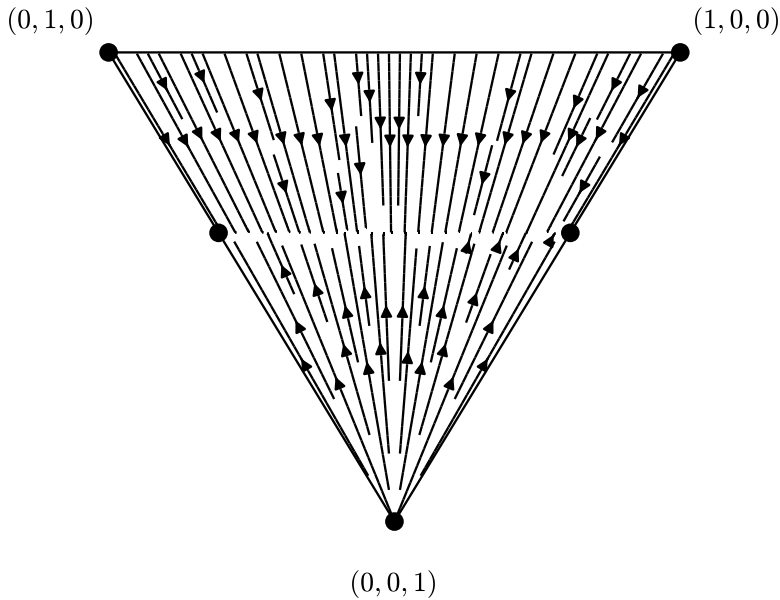

FIG. 1. **Example of hyperplanes in the phase space.** Competitors 1 and 2 (right and left corners respectively) are given the strategy $p = 1/2$, and the strategy of competitor 3 (bottom corner) is set to $p = 1/5$. The matrix $\boldsymbol{P}$ is not full rank, and the flow is therefore degenerated—the only interactions arise from the presence of competitor 3. A line of fixed points appears parallel to the (1,0,0)–(0,1,0) edge of the simplex.

## V. COMPLETE TWITTER ANALYSIS AND MODELING

### A. Raw data from Ref. [13] in main text.

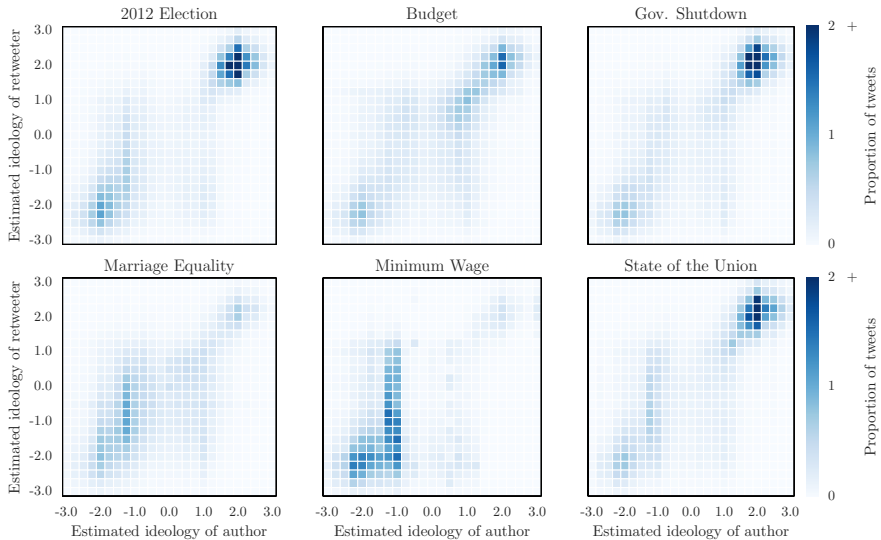

FIG. 2. **Political tweets**. Proportion of tweets by authors of estimated ideology $x$ retweeted by users of estimated ideology $y$, where $x, y \in [-3, 3]$ denotes the estimated ideology of the users. Strongly liberal users are given a score of $-3$ and strongly conservative users are given a score of $+3$. Topics are, from left to right and top to bottom: 2012 election, federal budget, government shutdown, marriage equality, minimum wage and the state of the union.

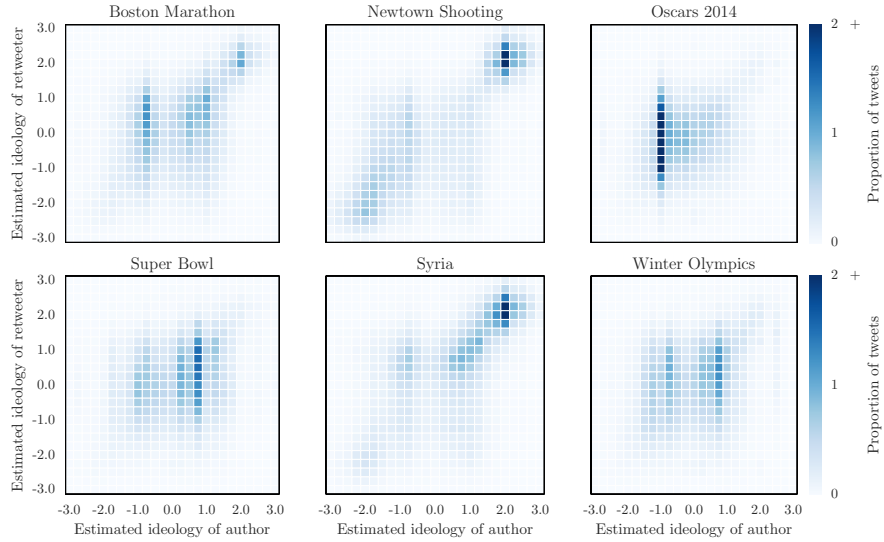

FIG. 3. **Non-political tweets**. Proportion of tweets by authors of estimated ideology $x$ retweeted by users of estimated ideology $y$, where $x, y \in [-3, 3]$ denotes the estimated ideology of the users. Strongly liberal users are given a score of $-3$ and strongly conservative users are given a score of $+3$. Topics are, from left to right and top to bottom: Boston marathon, Newtown shooting, 2014 Oscars, Super Bowl, Syria, and the 2014 Winter Olympics.

### B. Normalized matrices and coarse graining

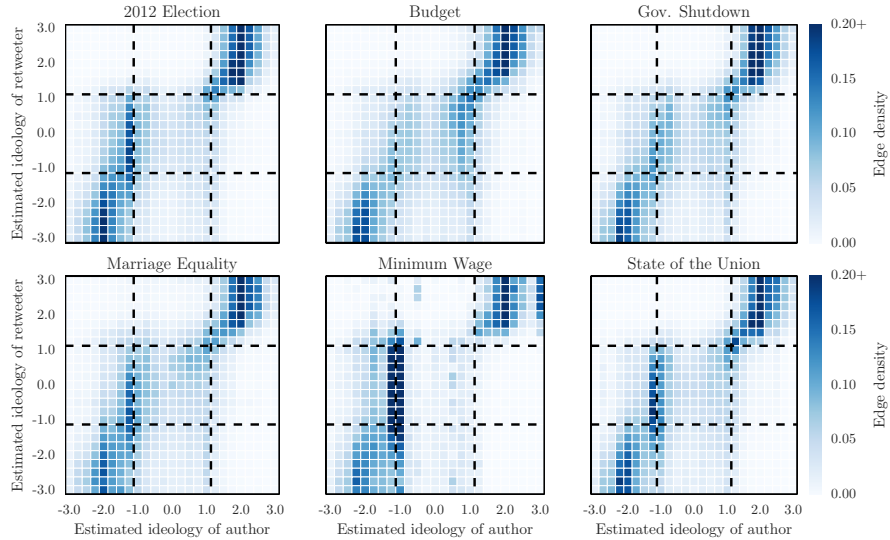

FIG. 4. **Strategies from political tweets.** Row-normalized retweet matrices yield potential strategy matrices. Dotted lines indicate the coarse-graining used to obtain a 3 competitor dynamics. Topics are, from left to right and top to bottom: 2012 election, federal budget, government shutdown, marriage equality, minimum wage and the state of the union.

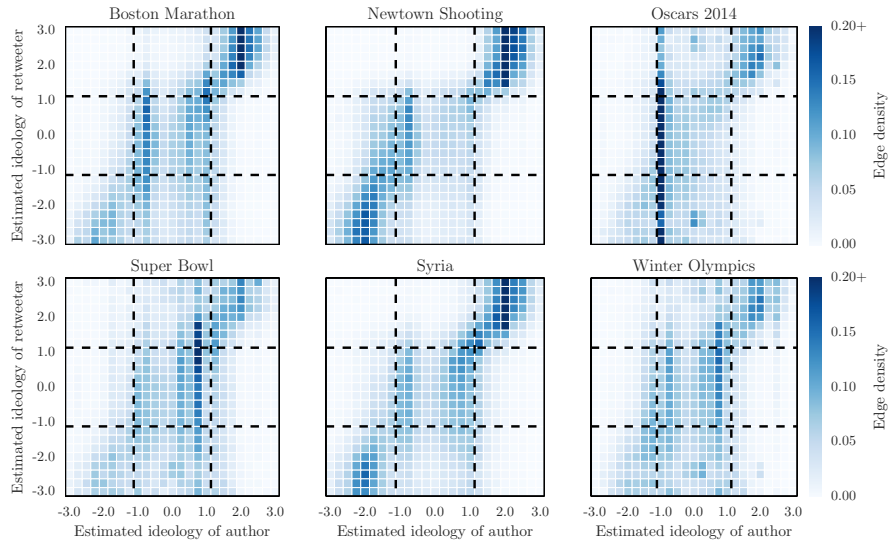

FIG. 5. **Strategies from non-political tweets.** Row-normalized retweet matrices yield potential strategy matrices. Dotted lines indicate the coarse-graining used to obtain a 3 competitor dynamics. Topics are, from left to right and top to bottom: Boston marathon, Newtown shooting, 2014 Oscars, Super Bowl, Syria, and the 2014 Winter Olympics.

### C. Flows in the prevalence space

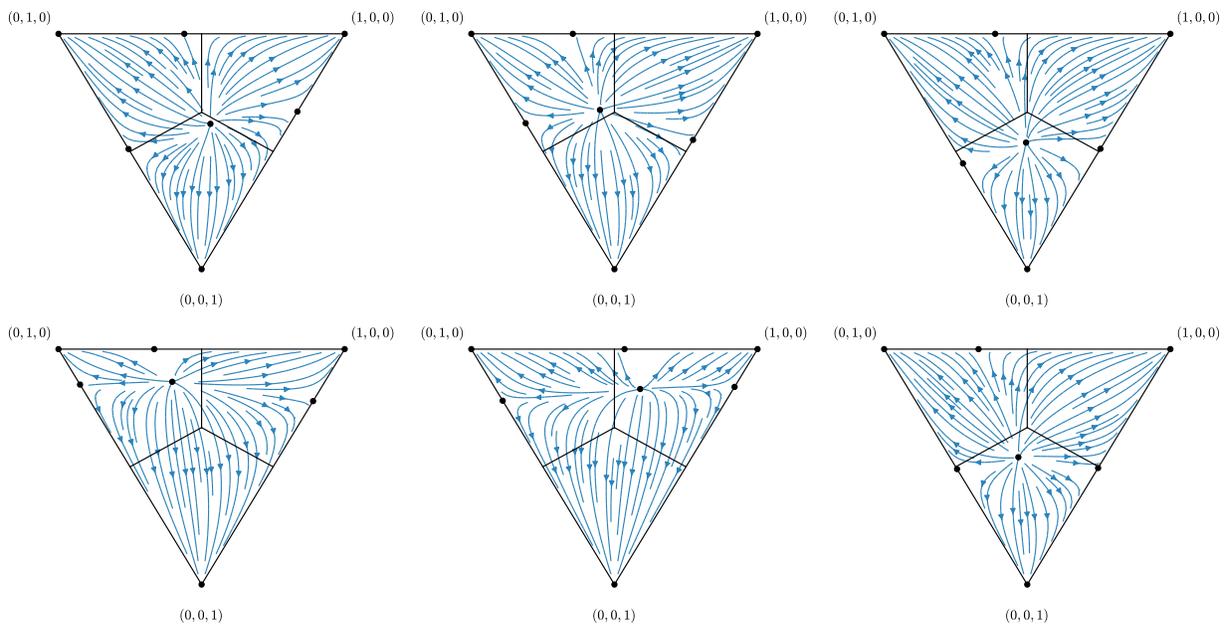

FIG. 6. **Flows in the prevalence space of political tweets.** The flows are obtained with our equations for arbitrary density matrices $\boldsymbol{P}$. Topics are, from left to right and top to bottom: 2012 election, federal budget, government shutdown, marriage equality, minimum wage and the state of the union. Point (1,0,0) correspond to a liberal outcome, (0,1,0) to a centrist outcome and (0,0,1) to a conservative outcome. All fixed points are shown with filled black circles, and their stability can be inferred from the trajectories around them.

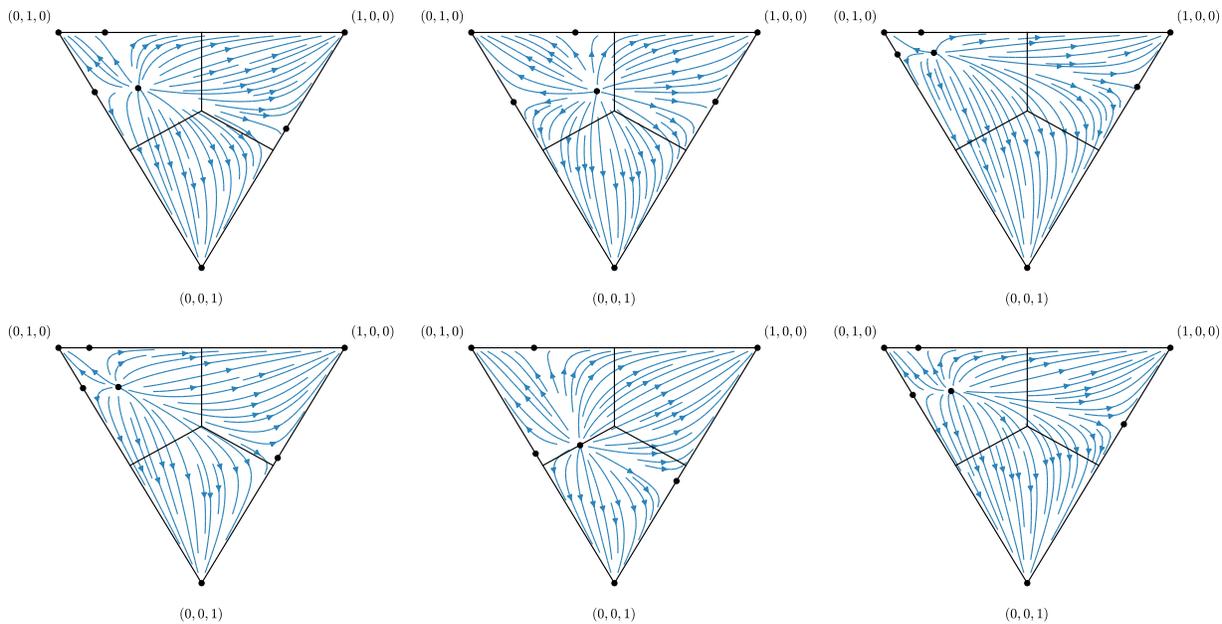

FIG. 7. **Flows in the prevalence space of non-political tweets.** The flows are obtained with our equations for arbitrary density matrices $\boldsymbol{P}$. Topics are, from left to right and top to bottom: Boston marathon, Newtown shooting, 2014 Oscars, Super Bowl, Syria, and the 2014 Winter Olympics. Point (1,0,0) correspond to a liberal outcome, (0,1,0) to a centrist outcome and (0,0,1) to a conservative outcome. All fixed points are shown with filled black circles, and their stability can be inferred from the trajectories around them.